\documentclass{elsart4-1}
\journal{C.R. Physique}
\volume{8}
\firstpage{385}
\lastpage{395}
\date{8 June 2007}
\pubyear{2007}
\runtitle{Y. Rabbia et al./ C.R. Physique 8 (2007) 385-395}
\runauthor{Y. Rabbia et al.}

\usepackage{graphicx}
\usepackage{amssymb}
\usepackage[english,francais]{babel}

\def\og{\leavevmode\raise.3ex\hbox{$\scriptscriptstyle\langle\!\langle$~}}
\def\fg{\leavevmode\raise.3ex\hbox{~$\!\scriptscriptstyle\,\rangle\!\rangle$}}

\begin{document}
\centerline{C.R. Physique 8 (2007) 385-395}
\vskip-2.5cm

\begin{frontmatter}

\selectlanguage{english}
\title{{\large\rm Optical techniques for direct imaging of exoplanets/Techniques
optiques pour l'imagerie directe des exoplan{\`e}tes} \\
The Achromatic Interfero Coronagraph}

\selectlanguage{english}
\author[authorlabel1]{Yves Rabbia\thanksref{CorrAuth}},
\thanks[CorrAuth]{Corresponding author}
\ead{\small\rm yves.rabbia@obs-azur.fr}
\author[authorlabel1]{Jean Gay},
\ead{\small\rm jean.gay@obs-azur.fr}
\author[authorlabel2]{Jean-Pierre Rivet}
\ead{\small\rm jean-pierre.rivet@obs-azur.fr}

\address[authorlabel1]{Observatoire de la C{\^o}te d'Azur,
 D{\'e}partement Gemini, UMR CNRS 6203, Av. Copernic, 06130 Grasse, France}
\address[authorlabel2]{Observatoire de la C{\^o}te d'Azur,
 D{\'e}partement Cassiop{\'e}e, UMR CNRS 6202, B.P. 4229, 06304 Nice cedex 04, France}

\smallskip
\begin{center}
{\small Available online 8 June 2007; DOI: 10.1016/j.crhy.2007.04.003}
\end{center}
\vskip-0.5cm

\begin{abstract}
\par
We report on the Achromatic Interfero Coronagraph, a focal imaging device which aims
at rejecting the energy contribution of a point-like source set on-axis, so as to
make detectable its angularly-close environment (applicable to stellar environment\,:
circumstellar matter, faint companions, planetary systems, but also conceivably
to Active Galatic Nucleii and multiple asteroids).

With AIC, starlight rejection is based on destructive interference, which allows
exploration of the star's neighbouring at angular resolution better than the
diffraction limit of the hosting telescope. Thanks to the focus crossing property
of light, rejection is achromatic thus yielding a large spectral bandwidth of work.
Descriptions and comments are given regarding the principle, the device itself,
the constraints and limitations, and the theoretical performance. Results are
presented which demonstrate the close-sensing capability and which show images
of a companion obtained in laboratory and \og on the sky\fg as well.
A short pictorial description of alternative AIC concepts, CIAXE and Open-Air CIAXE,
currently under study, is given.
{\it To cite this article: Y. Rabbia et al., C. R. Physique 8 (2007).}

\vskip 0.5\baselineskip

\selectlanguage{francais}
\noindent{\bf R{\'e}sum{\'e}}
\vskip 0.5\baselineskip
{\bf Le Coronographe Interf{\'e}rentiel Achromatique }
On pr{\'e}sente le Coronographe Interf{\'e}rentiel Achromatique (AIC dans le texte), un syst{\`e}me
imageur en mode coronographique pour rejeter de l'image, la contribution d'une source
ponctuelle sur l'axe de vis{\'e}e afin de laisser appara{\^\i}tre son environnement angulairement
proche (cela concerne les {\'e}toiles\,: mati{\`e}re circumstellaire, compagnons faibles, syst{\`e}mes
exoplan{\'e}taires, mais potentiellement aussi les Galaxies {\`a} noyau actif et les ast{\'e}ro{\"\i}des multiples).

Avec le AIC, la r{\'e}jection sur l'axe proc{\`e}de par interferences destructives, ce qui permet
une exploration du voisinage stellaire dans une proximit{\'e} angulaire meilleure que celle
fix{\'e}e par la limite de diffraction du t{\'e}lescope. Le principe de la r{\'e}jection utilise le
passage d'une onde par un foyer ce qui la rend achromatique et  permet ainsi d'observer
{\`a} large bande spectrale. On d{\'e}crit le principe et l'architecture du coronographe, les
contraintes instrumentales, les limitations associ{\'e}es, et les performances th{\'e}oriques
en r{\'e}jection. Des r{\'e}sultats de tests, en laboratoire et en observation sur le ciel, sont
donn{\'e}s, en terme d'images montrant les capacit{\'e}s de sondage proche et de d{\'e}tection d'un
compagnon faible. Une br{\`e}ve pr{\'e}sentation graphique de deux concepts CIAXE et Open-Air CIAXE
(actuellement en phase d'{\'e}tude) d{\'e}riv{\'e}s du principe g{\'e}n{\'e}rique est donn{\'e}e. 
{\it Pour citer cet article~: Y. Rabbia et al., C. R. Physique 8 (2007).}

\vskip 0.5\baselineskip
\noindent{\small{\it Mot-cl{\'e}s~:} Coronographie Stellaire; Imagerie {\`a} Haut Contraste;
Interf{\'e}rom{\'e}trie Annulante}}
\end{abstract}

\end{frontmatter}

\selectlanguage{english}

        \section{Introduction}
        \label{sec:Intro}

The Achromatic Interfero Coronagraph (AIC in the following) is a focal imaging
device working in coronagraphic mode and meant as a tool for the study of stellar
environment (Gay \& Rabbia~\cite{GR_CRAS_96}; Gay et al.~\cite{GR_CRAS_97};
Baudoz et al.~\cite{Baud_thes,Baud_a_AA1}). The purpose of  stellar coronagraphy
is to make detectable images of the faint emitting features located angularly very
close to an unresolved source whose brightness is tremendously larger (typically
$10^4$ to $10^6$ times larger) than any of the surrounding features. This is the
case for stars but also for other celestial sources as for example Active Galactic
Nuclei and asteroids with multiple components. In the following we use \og star\fg
and  \og companion\fg to respectively denote the central source and any neighbouring
feature. Any coronagraph aims at lowering as far as possible the energy contribution
of the parent star in the recorded image, which would otherwise prevent the
detection of the companion. The goal is thus to perform the \og extinction\fg
of the star while keeping the off-axis components transmitted as completely
as possible to the detector in image plane. This extinction capability, is usually
expressed by the \og Rejection rate\fg  defined by ${Rej} = F_{collected}/F_{residual}$,
where $F_{residual}$ and $F_{collected}$ are the fluxes recorded
with and without coronagraphic effect, respectively.

Not only the rejection capability is required but also the \og close-sensing\fg
capability\,: rejection must apply only within a very small angular extension,
called the \og Inner Working Angle\fg (IWA in the following) around the 
pointing direction, otherwise companions are also rejected.

In addition, the rejection ought to apply over as large as possible a spectral
bandwidth, benefiting to the detection capabilities in several respects\,:
increase of Signal to Noise Ratio (SNR), adaptability to science target by
flexible choice of dedicated spectral intervals, enhancement of detectivity
by a differential data processing, the \og adapted scale image subtraction\fg
(Gay et al.~\cite{Gay_subscaled}; Rabbia et al.~\cite{Rabbia_VLT})
inspired by a method introduced by Racine et al.~\cite{Racine}.

Early stellar coronagraphs have been a transposition to stars of the scheme
initially introduced by B.~Lyot (Lyot~\cite{Lyot}) to study the solar corona.
The technique relies on a tiny opaque mask set at the focus of an intermediate
image plane, so as to block photons from the on-axis source, while photons
from the companion are transmitted to the ultimate image plane. Because of
diffraction effects the mask must cover several Airy radii ($\lambda{/}D)$\,:
working wavelength/diameter of the telescope) what makes the \og close-sensing
capability\fg  comparatively poor (Mouillet et al.~\cite{DMouillet};
Beuzit et al.~\cite{JLBeuz}).

Another approach, called interfero-coronagraphy, yields the extinction from
a destructive interference process. The collected incident wave is splitted
into two components, which are recombined after insertion of a $\pi$ phase
shift between them. Thus the recombination (coherent addition of fields at
the detector) results in destructive interference and ideally no photon
from the on-axis source can reach the detector. Here, thanks to the
coherence properties of light, an angular sensing capability at a level
better than the diffraction limit ($\lambda/D$), is achievable. The
subsequent counterpart is a limitation of extinction (finite star size
and random tilt of incident wavefront).

As early as 1996, the concept of the AIC has been suggested 
(Gay \& Rabbia~\cite{GR_CRAS_96}; Gay et al.~\cite{GR_CRAS_97}) from which
various prototypes have been developed and used for tests in laboratory 
(Baudoz \cite{Baud_thes}; Rivet et al.~\cite{Rivet_iau200}) and  \og on the sky\fg
(Baudoz et al.~\cite{Baud_a_AA1,Baud_b_AA2}) on ground-based telescopes (although AIC
 has been initially devised for space-based operation
(Rabbia et al.~\cite{Rabbia_NGST})). Besides, even although coronagraphs are generally
meant as working with single compact aperture, AIC can be used for nulling
interferometry (Gay \& Rabbia~\cite{GR_CRAS_96}; Rabbia et al.~\cite{Rabbia_DomeC}),
for example, in a two-aperture Bracewell configuration (Bracewell~\cite{Bracwl_exopl}).
The immediate and essential requirement to obtain the coronagraphic effect with AIC
is that the (complex) transmission of the aperture must be centro-symmetric
(insensitive to a $180$~degrees rotation).

Other concepts for interfero-coronagraphy, each using a specific type of beam
separation/recombination have been devised since. Let us cite among other early
concepts\,:  the Phase Mask Coronagraph -PMC- (Roddier \& Roddier~\cite{Roddier_PMC}),
the Sectorized Phase Mask Coronagraph -SPMC- (Rouan et al.~\cite{Rouan_SMC}),
and the Phase Knife Coronagraph -PKC- (Abe et al.~\cite{Lyu_PKC}).

The main specific feature and advantage of AIC is the achromaticity of the phase
shift process, which was not the case for the others cited, early interfero-coronagraph
designs, and for the Lyot configuration as well. Another one regards the IWA, with
the best close-sensing capability (Guyon et al.~\cite{Guyon}). Another specific
feature is that AIC yields two twin-images of the companion (displayed symmetrically
with respect to the pointing axis), each conveying $1/4$ of  the collected energy.
This can be seen as a drawback because of a reduced transmission and because of
the symmetrization of companions (point-like or extended). Actually this is an
advantage for point-like companions whose separations and orbits are better
determined from the twin-images. 

In this article we give but a kind of status report on the AIC (a comparison between
AIC and other coronagraphs is given in Guyon et al.~\cite{Guyon}) focusing, for the
reader's convenience, on the underlying algebraic formalism, though recalling only
basic lines because of space limitation. In Section~\ref{sec:Basis} we describe the
principle of the device and the device itself and we give a short pictorial
description of alternative AIC devices, currently under study
(Gay et al.~\cite{Gay_ciaxe_cras,Gay_ciaxe_iau200}). In Section~\ref{sec:TechConst}
we describe the constraints for both implementation and operation, and we outline
the theoretically expected capabilities of AIC. In section~\ref{sec:Results}
we give results from testing AIC both in laboratory and \og on the sky\fg.

        \section{Basic principle}
        \label{sec:Basis}

    \subsection{Functional description and pictorial summary}
    \label{ssec:FuncrDesc}
    
\begin{figure}[h]
  \centering \resizebox{0.8\hsize}{!}{\includegraphics{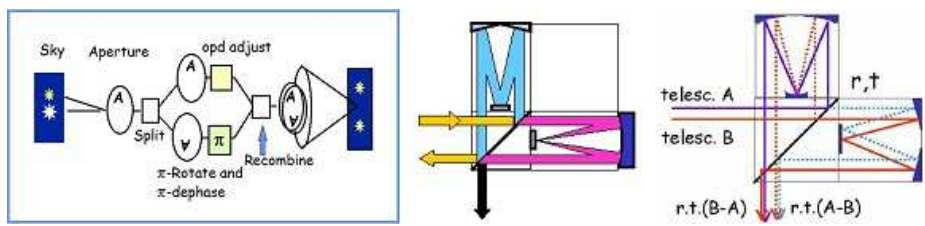}} 
  \caption{Schematic illustration of AIC functions (left) the configuration
  of beams in single aperture mode (center) )the routing of beams with a
  two-aperture interferometer (right).}
  \label{fig:cia_functions_and_beams}
\end{figure}

\begin{figure}[h]
  \centering \resizebox{0.9\hsize}{!}{\includegraphics{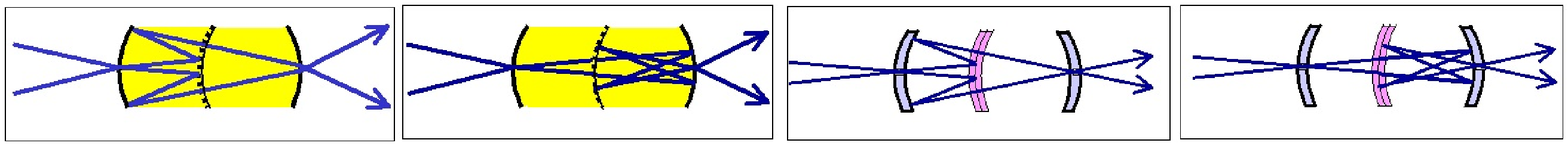}} 
  \caption{Schematic description with separated beam routes for CIAXE (left) and
  for Open-Air CIAXE (right), where generic functions of AIC appear. In CIAXE,
  beam entry and exit are small uncoated area, while they are holes in Open-Air
  CIAXE. Typical overall volume roughly is the one of a cylinder $4\times3$~mm
  length$\times$diameter.}
  \label{fig:ciaxe_horizontal}
\end{figure}

Basically, AIC is a Michelson-Fourier interferometer modified by inserting on
one arm an achromatic $\pi$ phase shift and a pupil rotation by $180$~degrees.
This double operation is performed by a cat's eye optical system, where the
$\pi$ phase shift originates in the focus crossing property (Gouy~\cite{Gouy};
Born \& Wolf~\cite{BW}; Boyd~\cite{Boyd}). The collimated beam from the
telescope is splitted in two sub-beams forming the two interferometric arms,
one (the \og fc\fg arm), where the focus-crossing occurs, the other
(the \og ff\fg arm) which includes a train of flat mirrors to balance the
optical paths in the interferometer. In the design these optical paths are
equal but some spurious residual Optical Path Difference (OPD) might still
occur at the implementation and must be controlled. The beamsplitter
at entry is used also for the recombination of beams. As a pictorial summary,
Fig.~\ref{fig:cia_functions_and_beams} illustrates these comments.
Mono-axial configurations, easier to insert in a telescope's optical train,
(Gay et al.~\cite{Gay_ciaxe_cras,Gay_ciaxe_iau200}) have been devised, and
are currently under study\,: the CIAXE and the open-air CIAXE.
Fig.~\ref{fig:ciaxe_horizontal} schematically illustrates the paths of beams
for the two concepts, and shows how the generic functions (beamsplitting,
focus-crossing, recombination, zero-OPD) are performed.

    \subsection{Physical description}
    \label{ssec:PhysDesc}
    
The typical AIC device consists of a kit of optical components assembled on a
basement comprising two galleries for beams to travel in air. All elements are
in Silica and assembling is made using molecular bonding under interferometric
control. Optical paths are equalized by construction, however a fine-tuning
of OPD (few nanometers) remains possible. The K photometric window,
($\lambda = 2.2\,\mu$m, $\Delta\lambda = 0.4\,\mu$m) being a convenient spectral
interval (regarding both science and constraints), the beamsplitter cube is
optimized for $\lambda = 2.2\,\mu$m but can work with a bandwidth larger than
the K-band. The remaining components being mirrors are achromatic by nature.
\begin{figure}[h]
  \centering \resizebox{1.0\hsize}{!}{\includegraphics{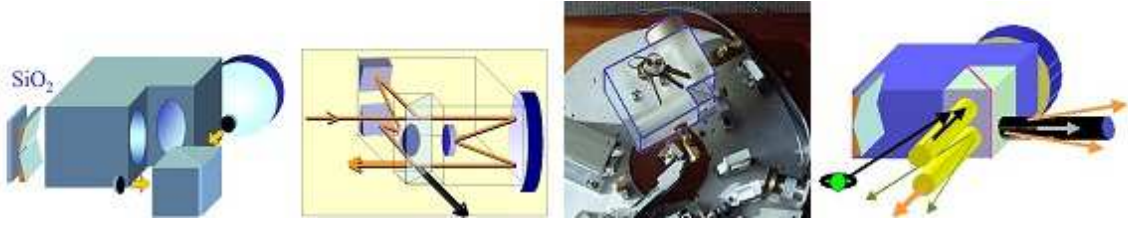}} 
  \caption{ From left to right\,: the kit of elements, the life of an on-axis
  beam and the two exit ports, AIC itself in an optical interface for accommodation
  on a telescope, the making of the output beams from both on-axis and off-axis sources.}
  \label{fig:cia_4_vues }
\end{figure}
An incident collimated on-axis beam yields two perpendicular output beams.
One, the constructive port, conveys all collected energy while no photon
goes into the other, the destructive port. An off-axis source yields two
symmetrically oriented sub-beams at each exit port. Each exiting sub-beam
conveys $1/4$ of the energy collected from the companion.

    \subsection{Formalism and algebraic description of the principle}
    \label{ssec:Formalism}
    
The formalism is the one of Fourier optics (Goodman \cite{Goodman}),
involving a pupil plane and an image plane. The cat's eye being designed
in this purpose, the two arms share the same pupil plane, placed at the
exit collimator, which focuses the recombined beams to the detector set
in the image plane.

\begin{figure}[h]
  \centering \resizebox{0.8\hsize}{!}{\includegraphics{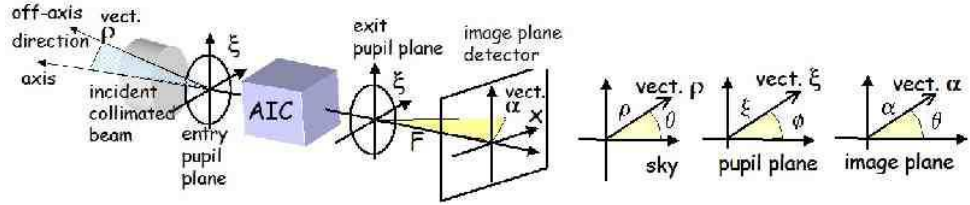}} 
  \caption{Schematic description of the notations used.}
  \label{fig:notations_vecteurs}
\end{figure}

Coordinates are vectorial eventhough we frequently note them by a single letter.
We use their reduced form (dimensionless). Thus, coordinate in pupil plane is
noted $\xi$ and means (vector\ in\ pupil\ plane){/}(wavelength)  while $\alpha$
is angular (radians) and means (vector\ in\ image\ plane){/}(focal\ length\ of
exit\ collimator). The magnification factor between entry and output pupil
planes is set as unity thus $\alpha$ is associated to an angle over the sky,
frequently noted $\rho$ (vector). All vector coordinates are sometimes
expressed using polar coordinates ($\xi$,$\phi$) and ($\alpha$,$\theta$)
or ($\rho$,$\theta$) where $\xi$, $\rho$ and $\alpha$ are then the moduli 
whilst $\phi$ and $\theta$ are the polar angles. Fig.~\ref{fig:notations_vecteurs}
illustrates the notations used.
Complex amplitudes of fields are noted $\psi_{collected}$ and $\psi_{recomb}$
respectively at the entry and at the exit pupil planes. The transmission $P(\xi)$
at the aperture is complex. Reflection and transmission coefficients for complex
amplitude at the beamsplitter are respectively noted $r$ and $t$, with as usual\,:
$R = \left|r\right|^{2}$  and $T = \left|t\right|^{2}$. Ideally $R=T=0.5$.

\subsubsection{Destructive interference process and off-axis images}
\label{sssec:DesyInt}
The complex amplitude of the field entering AIC is  $\psi_{collected}(\xi)= 
A.P(\xi).e^{i.\varphi(\xi)}$ since some phase distorsions $\varphi(\xi)$
might exist in the  collected wavefront. The field amplitude $A$ is such
that $\left|A\right|^{2} = \Omega$, the brightness of the on-axis source. 
At recombination, each sub-field have been both transmitted and reflected
at the beamsplitter. On its way each field might experience phase
distorsions (optical defects) and conveys the  phase variation caused
by propagation along the arm. In addition an extra optical path (yielding
$\varphi_{opd}$) can be inserted and eventually fine-tuned in the
\og ff\fg arm. Besides, in the \og fc\fg arm the $\pi$ phase shift occurs
and $\psi_{collected}(\xi)$ is symmetrized by the $180$~degrees rotation
of pupil, thus yielding $\psi_{collected}(-\xi)$. At recombination we
have $\psi_{recomb}(\xi)= \psi_{ff}(\xi)+\psi_{fc}(\xi)$ whose explicit
expression depends on the exit port. 
At the destructive port we have\,:
\begin{displaymath}
  \psi_{ff}(\xi)= r.t.A.P(\xi).\left[ 
  e^{i.\varphi(\xi)}.e^{i.\varphi_{ff}(\xi)}.
  e^{i.\varphi_{ff\_path}}.e^{i.\varphi_{opd}}\right]
  \end{displaymath}
and\,:
\begin{displaymath}
  \psi_{fc}(\xi)= r.t.A.P(-\xi).\left[e^{i.\pi}.
  e^{i.\varphi(-\xi)}.e^{i.\varphi_{fc}(\xi)}.
  e^{i.\varphi_{fc\_path}}\right].
\end{displaymath}
At the constructive port, $\psi_{ff}(\xi)$ keeps the same expression,
while the $\pi$ phase shift factor $e^{i.\pi}$ is replaced by unity
in $\psi_{fc}(\xi)$. In these description $\varphi_{ff}$ and $\varphi_{fc}$
are gathering internal phase defects in respective arms.
All phase terms are chromatic, but for easier reading,
$\lambda$ is omitted in the formulae.

\par{\it In ideal situation}, all mentioned phases are zero and optical
paths are equal, hence we have\,:
\begin{displaymath}
  \psi_{recomb}(\xi)= r.t.A.\left[ P(\xi)-P(-\xi) \right].
\end{displaymath}
Thus, as soon as the complex transmission is centro-symmetric ($P(\xi)=P(-\xi)$),
we find that $\psi_{recomb}(\xi)$ is uniformly zero. Hence photons from the
on-axis source do not reach the detector's plane (they are sent back to the
sky via the constructive port). This complete cleaning of the image plane for
stellar photons in ideal case is specific of AIC. Therefore, there is no
geometrical link between the rejection lobe (IWA) and the energy
distribution in image plane.

For a companion, off-axis by $\rho$, we have\,:
\begin{displaymath}
  \psi_{collected}(\xi)= A.P(\xi).e^{ - i.2.\pi.\xi.\rho},
\end{displaymath}
and at recombination we find\,:
\begin{displaymath}
  \psi_{recomb}(\xi)= r.t.A.P(\xi)\left[ e^{ - i.2.\pi.\xi.\rho}
  - e^{ + i.2.\pi.\xi.\rho} \right],
\end{displaymath}
which is not zero, so that some light escapes the destructive process (let
us note that, without the pupil rotation, we would find zero just like for
the on-axis source). Besides, the system yields two twin-images of the
companion, as described below, using Fourier Optics. The intensity in image
plane is\,:
\begin{displaymath}
  I(\alpha,\rho) = \left|\widehat{\psi}_{recomb}(\alpha)\right|^{2}.
\end{displaymath}
Since we have\,:
\begin{displaymath}
  \widehat{\psi}_{recomb}(\alpha) =
   r.t.\widehat{P}(\alpha)\ast\left[(\delta(\alpha - \rho)- 
   \delta(\alpha + \rho)\right],
\end{displaymath}
we end up (usual complex notations) with\,:
\begin{displaymath}
  I(\alpha,\rho) = R.T.\Omega.\left[\left|\widehat{P}(\alpha - \rho)\right|^{2}
  + \left|\widehat{P}(\alpha + \rho) \right|^{2} - 2.Re(\widehat{P}(\alpha - \rho)
  .\widehat{P}^{*}(\alpha + \rho))\right],
\end{displaymath}
that shows the presence of two twin images (centro-symetrically located Airy patterns)
and of a cross-term. This latter is vanishing as $\rho$ is increasing (enlarged
separation, product of amplitudes null). Conversely, it makes the intensity
progressively cancel as $\rho$ goes to zero (companion no longer off-axis,
hence undergoing rejection). We report in Fig.~\ref{fig:spatial_resp_wavefronts}
a pictorial summary of the destructive interference process both for star and
companion, in terms of both wavefronts and complex amplitudes.

\par{\it When departing from the ideal situation}, all mentioned phases no
longer are zero, but we still assume $P(\xi) = P(-\xi)$, the essential
requirement for the coronagraph to work properly. The amplitude at recombination
now is (with self-explanatory notations)\,:
\begin{displaymath}
  \psi_{recomb}(\xi)= r.t.A.P(\xi)\left[ e^{i.\varphi_{1}(\xi)}-
  e^{i.\varphi_{2}(\xi)} \right],
\end{displaymath}
which is clearly not zero, so that a residual energy is found in image plane
and rejection has a finite value, whose expression we now derive.

\subsubsection{Expression of the rejection from integrated energy in image plane} 
\label{sssec:RejIntEner}
In the introduction, rejection has been defined by the ratio $Rej = w_{0}/w$
with $w$  and $w_{0}$ the intensities recorded respectively with and without
the coronagraphic effect (in other words at destructive and constructive ports).
With this definition we have in real situation\,:
\begin{displaymath}
  Rej = \frac{w_{0}}{w}= \frac{\int \left|P(\xi).e^{i.\varphi_{1}(\xi)}
  + P(-\xi).e^{i.\varphi_{2}(\xi)}\right|^{2} d\xi}{\int\left|P(\xi).
  e^{i.\varphi_{1}(\xi)}- P(-\xi).e^{i.\varphi_{2}(\xi)}\right|^{2} d\xi}.
\end{displaymath}  
The phase factor that might occur in $P(\xi)$ is reported in the incident
wavefront phase distorsions, so that $P(\xi)$ is real and in addition
we assume\,
\begin{displaymath}
  P(\xi)=\Pi(\frac{\xi}{D})= \left|P(\xi)\right|^{2}
\end{displaymath}
to be the usual \og camembert-like\fg  transmission ($1$ inside, $0$ outside
the disk of diameter $D$). Moreover, it is reasonable to assume that the
various phase terms are small, so that using  the approximation
$e^{i.x} = 1 + i.x +....$ we write\,:
\begin{displaymath}
  Rej = \frac{\int_{pupil}(1 + cos\Delta\varphi(\xi)).
  d\xi}{\int_{pupil}(1 - cos\Delta\varphi(\xi)).d\xi}
  \approx \frac{4}{\int_{pupil}\left|\Delta\varphi(\xi)\right|^{2}
  d\xi} =\frac{4}{\sigma^{2}_{\Delta\varphi}},
\end{displaymath}
where we have introduced the differential phase $\Delta\varphi(\xi)$ such that\,:
\begin{displaymath} 
  \Delta\varphi(\xi)= \left[\varphi(\xi)- \varphi( -\xi)\right]
  + \left[\varphi_{ff}(\xi)- \varphi_{fc}( \xi)\right]
  + \left[\varphi_{ff\_path}- \varphi_{fc\_path} +\varphi_{opd}\right],
\end{displaymath}
and where $\varphi_{opd}$ allows adjusting to zero the mean-value, hence
the use of the variance $\sigma^{2}_{\Delta\varphi}$.
Respective expressions of integrated energy at constructive and destructive
ports are $w_{0} = 4.R.T.\Omega.S$ and
$w = \left[w_{0}{/}4\right].\sigma^{2}_{\Delta\varphi}$,
where $S$ is the area of the collecting aperture.

From the generic expression ${Rej}=4/\sigma^{2}_{\Delta\varphi}$,
the rejection can be evaluated in connection with the various phase
defects and conversely, technical specifications (tolerancing)
can be defined according to a targeted rejection; this is considered
later in the text.

Let us note that, although this definition is a general and widespread convention,
it rather traces the quality of the set-up than the ability to detect a companion,
since it does not consider the shape of the energy distribution in the image plane.
Therefore, for a companion imaged at a location free from residual energy,
this integrated quantity is not appropriate and likely to be pessimistic.
This point is also discussed later in the text.

\subsubsection{Spatial response of AIC\,: close-sensing}
\label{sssec:SpatResp}
To evaluate the close-sensing capability (in other words, beyond which angular
separation can the companion be seen?) we consider a point-like source off-axis
by $\rho$, and we integrate the energy in the image plane to obtain the recordable
energy with respect to $\rho$, what describes the spatial response $w(\rho)$
of AIC ($w(\rho)$ has a circular symmetry).

\begin{figure}[h]
  \centering \resizebox{0.9\hsize}{!}{\includegraphics{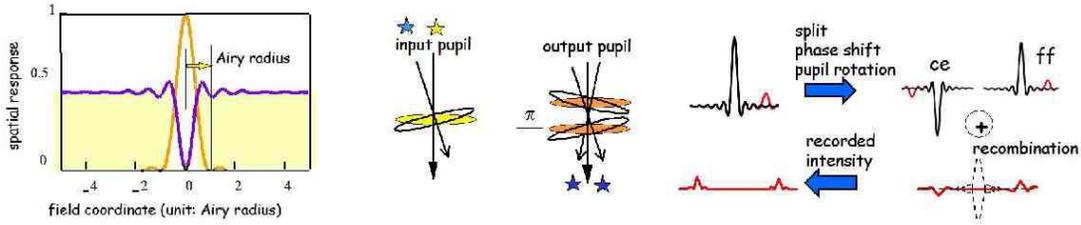}} 
  \caption{Left\,: radial profile of the spatial response of AIC (circular symmetry);
  Center and right\,: pictorial description of the destructive interference process,
  wavefront behaviour, amplitude behaviour.}
  \label{fig:spatial_resp_wavefronts}
\end{figure}

With our notations the brightness distribution (using Dirac $\delta$ symbol)
is $\Omega(\alpha,\rho) = \Omega.\delta(\alpha - \rho)$. Here we momentarily
drop all spurious phase effects, including OPD misbalance;
hence $\Delta\varphi(\xi) = 0$. We have then\,
\begin{displaymath}
  \psi_{collected}= A. P(\xi).e^{- i.2\pi.\xi.\rho},
\end{displaymath}
and using the Parseval-Plancherel theorem we can write\,:
\begin{displaymath}
  w(\rho)= \int I(\alpha,\rho).d\alpha =
  \int\left|\widehat{\psi}_{recomb}(\alpha,\rho)\right|^{2}.d\alpha =
  \int\left|\psi_{recomb}(\xi,\rho)\right|^{2}.d\xi.
\end{displaymath}
More explicitly, using polar coordinates ($\xi$,$\phi$) in pupil plane and 
($\rho$ ,$\theta$) for the off-axis direction, we write\,:
\begin{displaymath}
  w(\rho)= 2.R.T.\Omega.\int^{2.\pi}_{0}\int^{D/2}_{0}( 1 - e^{-i.2.\pi.(2.\xi.\rho).
  \cos{(\phi-\theta)}}.\xi.d\xi.d\phi.
\end{displaymath}
Thus, from Hankel's transform properties (Bracewell \cite{Bracwl_FT})
we end up with the spatial transmission or rejection lobe (circular symmetry,
sky coordinate $\rho$)\,:
\begin{displaymath}
  w(\rho)= 2.R.T.\Omega.(\frac{\pi.D^{2}}{4}).
  \left[ 1 - \frac{2.J_{1}( 2.\pi.D.\rho)}{2.\pi.D.\rho}\right],
\end{displaymath}
where the Airy distribution (for amplitudes)appears, associated to a telescope of
diameter $2D$. Fig.~\ref{fig:spatial_resp_wavefronts} illustrates the radial
profile of the function $w(\rho$), showing the close-sensing capability of AIC
and how it beats the diffraction-limit. Namely the maximal transmission occurs
at less than the telescope's first Airy ($\lambda/D$) ring. For an off-axis
source at $\rho$ = $0.3$~Airy, the transmission remains $0.5$ of that
far from axis (Baudoz \cite{Baud_thes}). 

        \section{Operational and technical constraints, associated limitations and theoretical capabilities}
        \label{sec:TechConst}

For the coronagraphic effect to work, two immediate constraints apply\,: 
(i) the complex transmission at exit pupil plane must be insensitive to a
$180$~degrees rotation, and the optical path difference (OPD) must accurately
be zero. Other constraints concern the technical specifications and
the conditions of observation.

\par
\begin{figure}[h]
  \centering \resizebox{0.9\hsize}{!}{\includegraphics{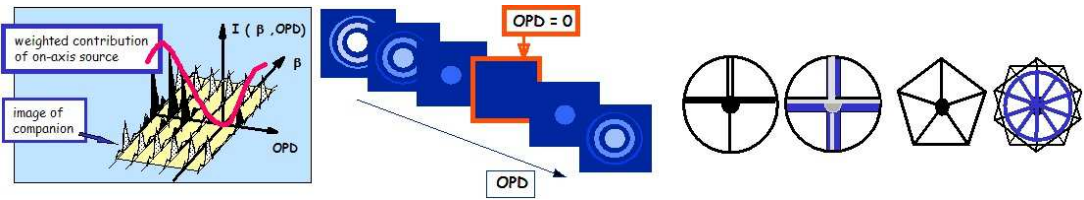}} 
  \caption{Left and center\,: pictorial illustration of the role of OPD;
  right\,: recovering amplitude centro-symmetry in the collecting aperture,
  by using appropriate masks in intermediate pupil plane.}
  \label{fig:opd_et_masques}
\end{figure}

{\it Regarding the symmetry of the pupil}, problems might arise from the
amplitude of the transmission $P(\xi)$, because of the central obscuration
and the spider bearing it (central-symmetry would possibly broken). Then, by
inserting a suitably designed  mask in an auxiliary pupil plane, centro-symmetry
is recovered; thus this constraint is not a heavy one. We report in
Fig.~\ref{fig:opd_et_masques} an illustration of both the trouble and the
remedies. Problems regarding the phase of the complex transmission are by
far more serious and are considered later in the text.

\par{\it Residual Optical Path Difference (OPD) and optical defects.}
OPD between interfering waves works as a weight applied to the contribution of
the on-axis source. As OPD goes away from zero the on-axis source gradually
appears in the image, and this rapidly degrades the rejection. This is
schematically illustrated in Fig.~\ref{fig:opd_et_masques}. Either for a residual
OPD or for optical defects we use the formula $Rej = 4{/}\sigma^{2}_{\Delta\varphi}$
but in place of $\sigma^{2}_{\Delta\varphi}$, which makes sense only
for optical defects, we directly take $\left[4.\pi^{2}{/}\lambda^{2}\right].\delta^{2}$
where $\delta$ stands for either the residual OPD or for the standard-deviation
of surface defects over the pupil. Then, with a target rejection $G$ we define
the tolerance on $\delta$ via $Rej\geq G$ which yields the condition
$\delta \leq \lambda{/}(\pi.\sqrt{G})$. For example with $G = 10^{4}$
the constraint on $\delta$ is roughly $\lambda{/}300$ and for $G = 10^{5}$
it is $\lambda{/}1000$, which are rather stringent constraints so that
interferometric control is mandatory when assembling AIC. The same range
of specification applies, for example, to the beamsplitting cube, where
internal reflected waves must be kept destructively interfering; this
is a matter of well-controlled optical paths within the cube. Moreover,
phase defects occurring inside AIC cannot be corrected for, and degrade
the rejection. 

    \subsection{Limitations from conditions of observation}
    \label{ssec:LimCdts}
    
Basically, stellar leakage and wavefront phase distorsions are the immediate
causes limiting the rejection. Stellar leakage occurs with any nulling technique,
when the star has a finite angular diameter and is incompletely eliminated
by the spatial response because its profile around origin is not flat.
Wavefront distorsions originate in deterministic optical defects (arising
in the optical train) and random distorsions induced by atmospheric turbulence,
this latter causing the major limitation in ground-based observations
(with tip/tilt fluctuations being the heaviest contributor). Another
constraint occurs because of atmospheric refraction (beyond the scope of
this paper). With space-based observations, beside fine guidance residual
fluctuations (jitter) only faint internal phase defects are occurring
and are not that big a concern (Baudoz \cite{Baud_thes};
Rabbia et al. \cite{Rabbia_NGST}).

\par{\it Stellar leakage,} denoted $w(\Theta$) for a star whose brightness 
distribution is $\Omega(\rho)= \Omega.\Pi(\frac{\rho}{\Theta})$ with angular
diameter $\Theta$, is evaluated by the integral of the spatial response
$w(\rho)$ taken over the brightness distribution. Again from Hankel's
transform properties we have\,:
\begin{displaymath}
  w(\Theta)= \int^{2.\pi}_{0}\ \int^{\Theta/2}_{0}\ w(\rho).\rho.d\rho.d\theta =
  2.R.T.\Omega.\left[1-\frac{J_{0}(\pi.D.\Theta)}{\left(\pi.D.\Theta\right)^{2}/4}\right].
\end{displaymath}
As an example for $\Theta = 0.05$~Airy radius, stellar leakage (residual/collected) is $0.001$
(Baudoz et al. \cite{Baud_a_AA1}).

\par{\it Random phase distorsions from atmospheric turbulence.}
The spurious phase distribution is noted \,:
\begin{displaymath}
  \Delta\varphi(\rho,\theta) = \varphi(\rho,\theta) -
  \varphi(\rho,\theta + \pi),
\end{displaymath} usually expressed as an infinite weighted sum of Zernike
orthogonal polynomials\,:
\begin{displaymath}
  \varphi(\rho,\theta)= \sum^{\infty}_{j=1}w_{j}.Z_{j}(\rho,\theta),
\end{displaymath}
where the $w_{j}$ are random and trace the effect of turbulence. The resulting
variance $\sigma^{2}_{\Delta\varphi}$ must be reduced by using an Adaptive Optics (AO)
system. Its role is to induce appropriate phase distorsions so as to absorb some
of the incident ones. Note that centro-symmetric phase distorsions automatically
cancel, whilst other are amplified; however, a significant gain remains by
using AO (Baudoz et al. \cite{Baud_a_AA1}). For convenience, we refer to the
Zernike formalism (Noll \cite{Noll}) to theoretically describe the action
of AO, which leads to a corrected phase distribution\,:
\begin{displaymath}
  \varphi_{corr}(\rho,\theta)= \varphi(\rho,\theta)-
  \sum^{J}_{j=1}a_{j}.Z_{j}(\rho,\theta),
\end{displaymath} where $J$ is the highest Zernike order include in the correction.
The resulting variance is significantly less than the previous one, and hence
there is a better rejection. The result in the image is a centro-symmetric speckle
pattern, whose time-averaged distribution looks like a halo (volcano+caldera-shaped).
Algebraic derivation (Baudoz et al. \cite{Baud_a_AA1}) shows that the higher $J$,
the larger the caldera (central dark hole) and the lower the edges of the volcano.
Actually, this theoretical approach pertains to ultimate performance.
Practically, since AO may work in various regimes (zonal, modal, Fourier basis set)
the pure Zernike formalism might depart from real situations and expected corrections
are not completely efficient (what we call \og incomplete correction\fg )
and estimations must be reviewed, using an example of actual AO performance
(Conan \cite{Conan}). Both estimates from numerical simulation are shown in
Fig.~\ref{fig:theoretical_all}. 

\par{\it Effect of residual tip/tilt phase defects.} Incomplete tip/tilt
correction (pointing jitter) causes the heaviest degradation and deserves
a specific derivation. Taking into account that tip/tilt residuals are small,
and using the approximation $2.J_{1}(z)/z \cong 1 - z^{2}/8$ the resulting
average rejection is obtained from the averaged spatial response yielding\,:
\begin{displaymath}
  <Rej> = \frac{w_{0}}{<w(\rho>} \cong \frac{16}{\left(2.\pi.D/\lambda\right)^{2}.
  <\rho^{2}>},
\end{displaymath}
whence is extracted the constraint\,:
\begin{displaymath}
  \sigma_{\rho} \leq \frac{airy}{\sqrt{<Rej>}},
\end{displaymath}
where $<Rej>$ is the target rejection and $airy = 1.22 \frac{\lambda}{D}$.
For example, a $10^{4}$ target leads to $\sigma_{\rho} \leq \frac{airy}{100}$.
With HST ($D=2.4$~m) at $2.2 \mu$~m we find $\sigma_{\rho} \leq 10^{- 8}$
rad or $0.002$~arcsec, which is in the range of the HST fine guidance capability.

    \subsection{theoretically expectable performance}
    \label{ssec:TheoPerf}
    
\par
\begin{figure}[h]
  \centering \resizebox{0.6\hsize}{!}{\includegraphics{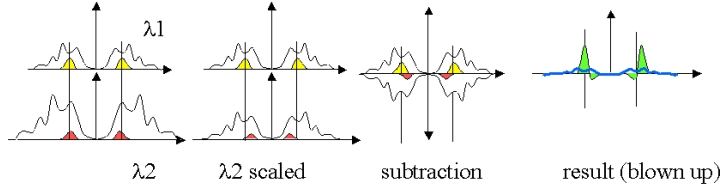}} 
  \caption{Pictorial summary of the scaled-subtraction process.}
  \label{fig:scaled_subtraction}
\end{figure}

{\it Enhancement of detection capability by the \og scaled-subtraction\fg process.}
Thanks to the large spectral bandwidth allowed by AIC, two adjacent channels splitting
the $K$ window might be accommodated on the same detector. The respective residual
speckles have homothetic patterns (scaling factor\,: $\lambda$). Therefore, in principle
(thanks to the central \og dark-hole\fg ) it is possible to give them the same spatial
distribution by properly scaling both the spatial extension and the intensity
distribution. Then, images of a companion become shifted in one channel with respect
to the other and have unequal heights. Subtracting the re-scaled patterns tends
to eliminate the speckles while a trace of the companion remains. This process 
(schematically summarized in Fig.~\ref{fig:scaled_subtraction}), is described
with numerical simulations in Gay et al. \cite{Gay_subscaled} and
Verinaud $\&$ Carbillet \cite{Verin_Carb_sub_scal}. In case of an on-axis residual
spurious contribution, efficiency is reduced, but a significant gain
in detection capability remains.

\par{\it Signal to Noise ratio for the detection of a companion.}
As already mentioned, using the integrated rejection ${Rej}= 4{/}\sigma^{2}_{\Delta\varphi}$
is not appropriate to evaluate the detection capability for a companion, the point being
not to count unwanted photons in the whole image plane, but rather to distinguish the
companion against the noisy and non-uniform residual energy distribution.
Thus the convenient approach for theoretical assessments is to express a pixel-dependent
Signal to Noise Ratio (SNR), taking into account the spatial response of AIC, the intensity
fluctuations (speckled halo) and the area covered by the companion's image
(distance to axis and image extension). From a detailed analysis
(Baudoz et al. \cite{Baud_a_AA1}) the expression of this SNR, ($M$ recorded exposures,
companion off-axis by angular vector $\rho$) is\,:
\begin{displaymath}
  SNR(\rho) =  
  \frac{(N_{c}/4).\sqrt{2}.\sqrt{M}.\int_{companion}w(\alpha).
  C(\alpha-\rho).d\alpha}{\sqrt{\int_{companion}\left[var_{residual}(\alpha)
  + var_{bg} + var_{detect}\right]. d\alpha}},
\end{displaymath}
where the factor $1/4$ traces the transmission for a single image, $\sqrt{2}$ 
accounts for the two twin-images, $w(\rho)$ is the spatial response of AIC,
and $C(\alpha-\rho)$ is the normalized distribution weighting of the
companion's contribution (Airy-like shaped). Integration is taken over the
pixels covered by the companion's image.
\begin{figure}[h]
  \centering \resizebox{0.9\hsize}{!}{\includegraphics{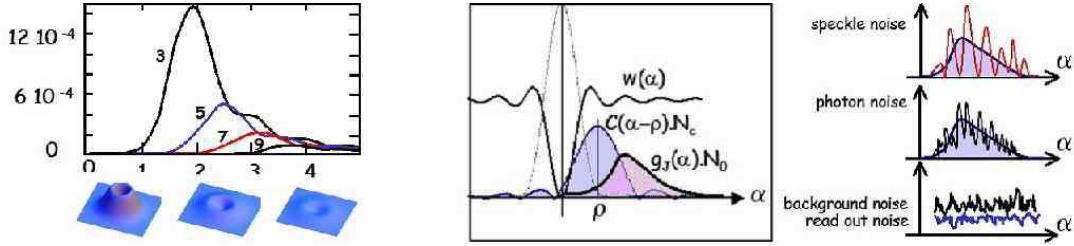}} 
  \caption{Left\,: residual energy profiles with AO corrections up to 9 Zernike radial orders
  (meaning $J = 36$); center\,: notations for the SNR expression (only one-sided cuts featured);
  right\,: phenomenology of noises in the residual energy.}
  \label{fig:halo_noise}
\end{figure}

The number of collected photons from a $m$-magnitude star is\,:
\begin{displaymath}
  N_{0}= \left[F_{ref}(\lambda){/}h.\nu\right].10^{ - 0.4\ m}.
  \left[\eta.4.R.T.S.t_{opt}.\Delta\lambda.\tau \right],
\end{displaymath}
with usual notations, while for the companion (magnitude difference $\Delta m$)
we have\,:
\begin{displaymath}
  N_{c} = N_{0}. 10^{ - 0.4\  \Delta m}.
\end{displaymath}
In the noise factor, variances are related to a given pixel\,: $var_{det}$
is the readout noise variance (ron$^{2}$) of the detector, $var_{bg}$ is for
the background radiation (using Planck's formulae)  and $var_{residual}$
comes from the fluctuations of the residual energy distribution. This latter
includes $g_{J}.N_{0}$  the residual energy when AO-corrections are carried
up to the Zernike radial order $J$ (each comprising several azimuthal orders),
and is evaluated by the doubly stochastic process involving the Poisson photon
noise at a given illumination level and the fluctuations of this level.
Performance of AIC (expressed as detectable $\Delta m$,
directly depends on AO capabilities (Fig.~\ref{fig:halo_noise}).
 
\begin{figure}[h]
  \centering \resizebox{0.9\hsize}{!}{\includegraphics{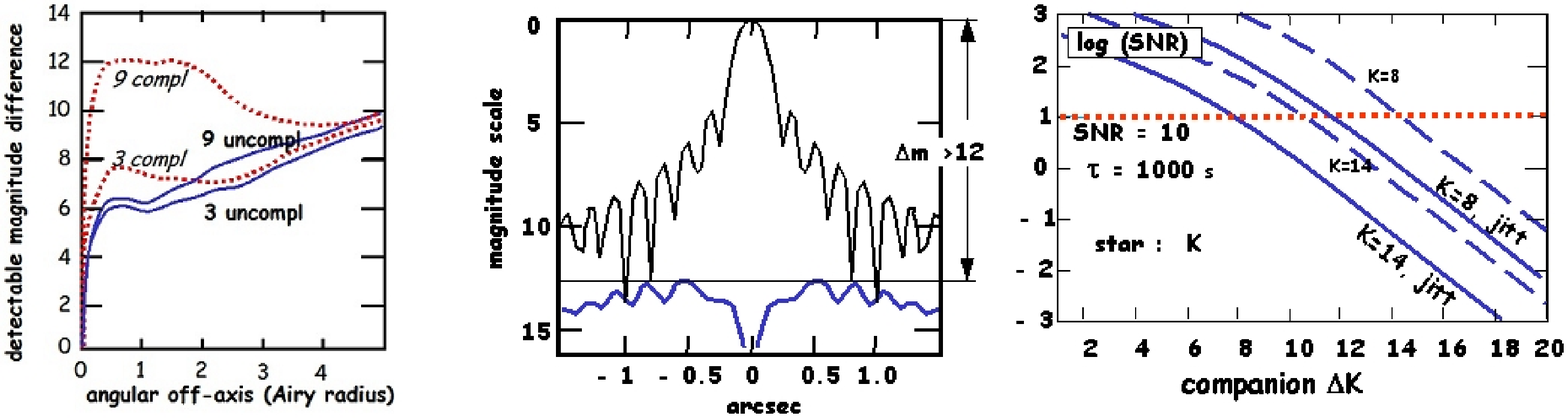}} 
  \caption{Left\,: theoretically detectable magnitude difference of a companion versus
  its off-axis angle, showing both complete (ideal) and uncomplete (real) AO corrections,
  for radial orders $3$ and $9$, calculated for expectable conditions at CFHT, Hawaii\,:
  $r_{0} = 30$~cm, $\tau = 0.1"$, SNR$=5$ for target $K=5$; (see Baudoz et 
  al.\cite{Baud_a_AA1,Baud_b_AA2}).
  Center\,: numerical simulation (based on a phase
  error map of HST) of residual energy profile for observation in K window ($2.2\mu$m;
  $\Delta\lambda = 0.4 \mu$m), and expressed in magnitude differences.
  Right\,:   behaviour of SNR versus $\Delta K$ the magnitude difference between star
  and companion; curves pertain to star K magnitudes $8$ and $14$. Dashed lines\,:
  optical quality is assumed the same as on HST, pointing instabilities eliminated.
  Solid lines\,: pointing residuals of $5$~mas rms causes less than 3 magnitudes
  degradation. K window, quantum efficiency $\eta = 0.5$, $R = T = 0.5$, $S = 50m^{2}$,
  optical transmission $t_{opt} = 0.1$, observing time $1000$~sec, ron$ = 20 e$/pix.}
  \label{fig:theoretical_all}
\end{figure}

Illustrations for \og complete\fg  and \og incomplete\fg  corrections
are given in Fig.~\ref{fig:theoretical_all} as well as expectable
detection capabilities in space-based observation.

        \section{Results}
        \label{sec:Results}

\begin{figure}[h]
  \centering \resizebox{0.9\hsize}{!}{\includegraphics{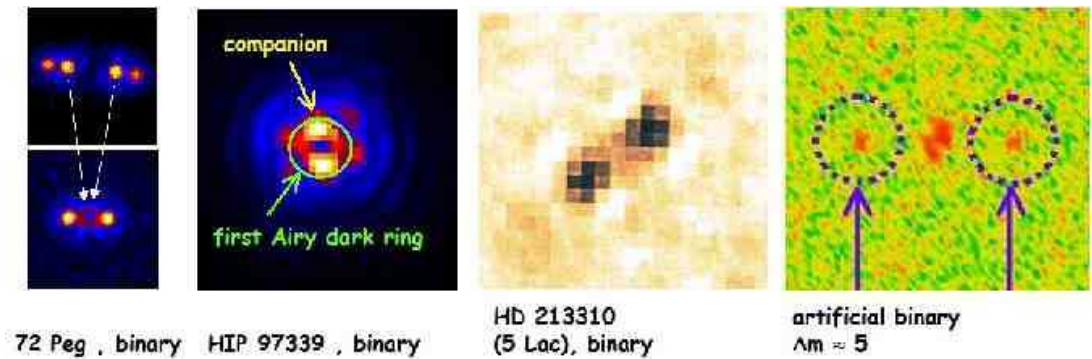}}
  \caption{Left, center left, center right\,: celestial sources, $\lambda = 2.2\mu$m,
  }
  \label{fig:results_all_ter}
\end{figure}

The results obtained so far both in laboratory and \og on the sky\fg are but
illustrations of potential capabilities. To spare space we rely on a pictorial
presentation, see Fig.~\ref{fig:results_all_ter} and
Baudoz et al.~\cite{Baud_a_AA1,Baud_b_AA2} for details.

Close-sensing capability is clearly validated. Effective capability for the detection
is in good agreement between expected and observed capabilities, taking into account
the turbulence conditions and the AO performance.

        \section{Conclusion}
        \label{sec:Concl}

In this article we have presented the main and specific features of our Achromatic
Interfero Coronagraph (principle, theoretical capabilities and some results from
test runs in laboratory and behind a telescope). The main advantages of AIC are
the large  spectral bandwidth over which the coronagraphic rejection is achievable
and the very small Inner Working Angle (providing the capability to explore the
very close angular neighbouring of a source at a level better than set by
the diffraction limit). The main drawback is the difficulty to achieve and to 
maintain equality of the optical paths at the required accuracy. Recently,
an improved version of the device (regarding OPD fine-tuning) has been built
and is presently under test. The future step is to accommodate this new AIC
on a large telescope equipped with Adaptive Optics, so as to undertake scientific
programs. New configurations, easier to accommodate in a telescope's
optical train are currently under study.

\end{document}